\documentclass[11pt]{article}

\pdfoutput=1

\usepackage[utf8]{inputenc}
\usepackage[T1]{fontenc}
\usepackage{hyphenat}
\usepackage{xspace}
\usepackage{amsmath}
\usepackage{amsfonts}
\usepackage{hyperref}
\usepackage{url}
\usepackage{booktabs}
\usepackage{multirow}
\usepackage{subfigure}
\usepackage{makecell}
\usepackage{caption}
\usepackage{minibox}
\usepackage{bbm}
\usepackage{graphicx}
\usepackage{balance}
\usepackage{mathtools}
\usepackage{color}
\usepackage{marvosym}
\usepackage{ifthen}
\usepackage{textcomp}
\usepackage{enumitem}
\usepackage{verbatim}
\usepackage{algorithm}
\usepackage{numprint}
\usepackage{balance}
\usepackage{lipsum}

\usepackage{booktabs}
\usepackage{dcolumn}
\usepackage{rotating}

\usepackage{amsthm}
\theoremstyle{plain}

\newcommand{\SupFig}[1]{Supplementary Fig.~#1}
\newcommand{\SupTab}[1]{Supplementary Table~#1}
\newcommand{\SupFigNumDocs}{\SupFig{1}}
\newcommand{\SupFigAltModels}{\SupFig{2}}
\newcommand{\SupFigCultVsCommMem}{\SupFig{3}}
\newcommand{\SupFigDocLength}{\SupFig{4}}
\newcommand{\SupFigCurveChars}{\SupFig{5}}
\newcommand{\SupFigCurveCharsByType}{\SupFig{6}}
\newcommand{\SupFigSilhouette}{\SupFig{7}}
\newcommand{\SupFigClusterOverlay}{\SupFig{8}}
\newcommand{\SupTabBioStats}{\SupTab{1}}
\newcommand{\SupTabCurveChars}{\SupTab{2}}
\newcommand{\SupTabTwitterCompleteness}{\SupTab{3}}

\newcommand{\N}{2\,362\xspace}
\renewcommand{\d}{\operatorname{d}}
\newcommand{\trackedChange}[1]{#1}
\newcommand{\trackedChangeNew}[1]{#1}

\newcommand{\chatoDisplayMode}[1]{#1}

\definecolor{MyRed}{rgb}{0.6,0.0,0.0} 
\definecolor{MyBlack}{rgb}{0.1,0.1,0.1} 
\newcommand{\inred}[1]{{\color{MyRed}\sf\textbf{\textsc{#1}}}}
\newcommand{\frameit}[2]{
  \begin{center}
  {\color{MyRed}
  \framebox[.9\columnwidth][l]{
    \begin{minipage}{.85\columnwidth}
    \inred{#1}: {\sf\color{MyBlack}#2}
    \end{minipage}
  }\\
  }
  \end{center}
}

\newcommand{\note}[2][]{\chatoDisplayMode{\def\@tmpsig{#1}\frameit{{\Pointinghand} Note}{#2\ifx \@tmpsig \@empty \else \mbox{ --\em #1}\fi}}}
\newcommand{\todo}[2][]{\chatoDisplayMode{\def\@tmpsig{#1}\frameit{{\Writinghand} To-do}{#2\ifx \@tmpsig \@empty \else \mbox{ --\em #1}\fi}}}

\newcommand{\abbrevStyle}[1]{#1}

\newcommand{\ie}{\abbrevStyle{i.e.}\xspace}
\newcommand{\eg}{\abbrevStyle{e.g.}\xspace}

\newcommand{\etAl}{\abbrevStyle{et al.}\xspace}
\newcommand{\vs}{\abbrevStyle{vs.}\xspace}
\newcommand{\etc}{\abbrevStyle{etc.}\xspace}

\newcommand{\Tabref}[1]{Table~\ref{#1}}
\newcommand{\Figref}[1]{Fig.~\ref{#1}}

\newcommand{\denselist}{ \itemsep -2pt\topsep-10pt\partopsep-10pt }

\newcommand{\textcite}[1]{\citeauthor{#1} \shortcite{#1}}

\newcommand{\hide}[1]{}

\DeclareMathOperator*{\argmin}{arg\,min}

\makeatletter
\newcommand{\iffont}[2]{\ifthenelse{\equal{\f@family}{#1}}{#2}{}}
\makeatother

\iffont{ptm}{
  \usepackage{mathptmx}

  \DeclareSymbolFont{greek}{OML}{cmm}{m}{n}
  \DeclareMathSymbol{\alpha}{\mathalpha}{greek}{"0B}
  \DeclareMathSymbol{\beta}{\mathalpha}{greek}{"0C}
  \DeclareMathSymbol{\gamma}{\mathalpha}{greek}{"0D}
  \DeclareMathSymbol{\delta}{\mathalpha}{greek}{"0E}
  \DeclareMathSymbol{\epsilon}{\mathalpha}{greek}{"0F}
  \DeclareMathSymbol{\zeta}{\mathalpha}{greek}{"10}
  \DeclareMathSymbol{\eta}{\mathalpha}{greek}{"11}
  \DeclareMathSymbol{\theta}{\mathalpha}{greek}{"12}
  \DeclareMathSymbol{\iota}{\mathalpha}{greek}{"13}
  \DeclareMathSymbol{\kappa}{\mathalpha}{greek}{"14}
  \DeclareMathSymbol{\lambda}{\mathalpha}{greek}{"15}
  \DeclareMathSymbol{\mu}{\mathalpha}{greek}{"16}
  \DeclareMathSymbol{\nu}{\mathalpha}{greek}{"17}
  \DeclareMathSymbol{\xi}{\mathalpha}{greek}{"18}
  \DeclareMathSymbol{\pi}{\mathalpha}{greek}{"19}
  \DeclareMathSymbol{\rho}{\mathalpha}{greek}{"1A}
  \DeclareMathSymbol{\sigma}{\mathalpha}{greek}{"1B}
  \DeclareMathSymbol{\tau}{\mathalpha}{greek}{"1C}
  \DeclareMathSymbol{\upsilon}{\mathalpha}{greek}{"1D}
  \DeclareMathSymbol{\phi}{\mathalpha}{greek}{"1E}
  \DeclareMathSymbol{\chi}{\mathalpha}{greek}{"1F}
  \DeclareMathSymbol{\psi}{\mathalpha}{greek}{"20}
  \DeclareMathSymbol{\omega}{\mathalpha}{greek}{"21}
  \DeclareMathSymbol{\varepsilon}{\mathalpha}{greek}{"22}
  \DeclareMathSymbol{\vartheta}{\mathalpha}{greek}{"23}
  \DeclareMathSymbol{\varpi}{\mathalpha}{greek}{"24}
  \DeclareMathSymbol{\varrho}{\mathalpha}{greek}{"25}
  \DeclareMathSymbol{\varsigma}{\mathalpha}{greek}{"26}
  \DeclareMathSymbol{\varphi}{\mathalpha}{greek}{"27}
  \DeclareSymbolFont{otone}{OT1}{cmr}{m}{n}
  \DeclareMathSymbol{\Gamma}{\mathalpha}{otone}{0}
  \DeclareMathSymbol{\Delta}{\mathalpha}{otone}{1}
  \DeclareMathSymbol{\Theta}{\mathalpha}{otone}{2}
  \DeclareMathSymbol{\Lambda}{\mathalpha}{otone}{3}
  \DeclareMathSymbol{\Xi}{\mathalpha}{otone}{4}
  \DeclareMathSymbol{\Pi}{\mathalpha}{otone}{5}
  \DeclareMathSymbol{\Sigma}{\mathalpha}{otone}{6}
  \DeclareMathSymbol{\Upsilon}{\mathalpha}{otone}{7}
  \DeclareMathSymbol{\Phi}{\mathalpha}{otone}{8}
  \DeclareMathSymbol{\Psi}{\mathalpha}{otone}{9}
  \DeclareMathSymbol{\Omega}{\mathalpha}{otone}{10}
  \DeclareSymbolFont{syms}{OML}{cmm}{m}{it}
  \DeclareMathSymbol{\partial}{\mathord}{syms}{"40}
  \DeclareMathAlphabet{\mathbold}{OML}{cmm}{b}{it}
  \DeclareSymbolFont{largesymbols}{OMX}{cmex}{m}{n}

}

\usepackage{authblk}
\usepackage{fullpage}
\usepackage{amssymb,amsmath}
\usepackage[left]{lineno}

\usepackage{setspace}

\title{Postmortem memory of public figures in news and social media%
\footnote{%
This manuscript was also published in
the \textit{Proceedings of the National Academy of Sciences,} September 2021, Vol.~118, No.~38, e2106152118,
available online (alongside Supplementary Information) at \url{https://doi.org/10.1073/pnas.2106152118}.
Please cite accordingly.
}%
}

\author[1$^\dagger$]{Robert West}
\author[2]{Jure Leskovec}
\author[3]{Christopher Potts}

\affil[1]{School of Computer and Communication Sciences,
EPFL,
1015 Lausanne, Switzerland}
\affil[2]{Department of Computer Science, Stanford University, Stanford, CA 94305, USA}
\affil[3]{Department of Linguistics, Stanford University, Stanford, CA 94305, USA}
\affil[$^\dagger$]{{\small To whom correspondence should be addressed: robert.west@epfl.ch}}

\date{}

\begin{document}

\maketitle

\begin{abstract}
\noindent
Deceased public figures are often said to live on in collective memory. We quantify this phenomenon by tracking mentions of 2\,362 public figures in English-language online news and social media (Twitter) one year before and after death. We measure the sharp spike and rapid decay of attention following death and model collective memory as a composition of communicative and cultural memory. Clustering reveals four patterns of post-mortem memory, and regression analysis shows that boosts in media attention are largest for pre-mortem popular anglophones who died a young, unnatural death; that long-term boosts are smallest for leaders and largest for artists; and that, while both the news and Twitter are triggered by young and unnatural deaths, the news additionally curates collective memory when old persons or leaders die. Overall, we illuminate the age-old question who is remembered by society, and the distinct roles of news and social media in collective memory formation.
\end{abstract}

\renewcommand{\abstractname}{Significance statement}
\begin{abstract}
\noindent
Who is remembered by society after they die? Although scholars as well as the broader public have speculated about this question since ancient times, we still lack a detailed understanding of the processes at work when a public figure dies and their media image solidifies and is committed to the collective memory. To close this gap, we leverage a comprehensive five-year dataset of online news and social media posts with millions of documents per day. By tracking mentions of thousands of public figures during the year following their death, we reveal and model the prototypical patterns and biographic correlates of post-mortem media attention, as well as systematic differences in how the news vs.\ social media remember deceased public figures.
\end{abstract}



\bigskip


\noindent
Being remembered after death has been an important concern for humans throughout history~\cite{carr_mortuary_1995},
and conversely, many cultures have considered \textit{damnatio memoriae}---being purposefully erased from the public's memory---one of the most severe punishments conceivable~\cite{flower_art_2006}.
To reason about the processes by which groups and societies remember and forget, the French philosopher and sociologist Maurice Halbwachs introduced the concept of collective memory in 1925~\cite{halbwachs_les_1925}, which has since been a subject of study in numerous disciplines, including anthropology, ethnography, philosophy, history, psychology, and sociology,
and which gave rise to the new discipline of memory studies~\cite{roediger_iii_creating_2008}.
Over the decades, collective memory has moved from being a purely theoretical construct to becoming a practical phenomenon that can be studied empirically~\cite{roediger_iii_collective_2015}, \eg, in order to quantify to what extent U.S.\ presidents are remembered across generations~\cite{roediger_iii_forgetting_2014} or how World War~II is remembered across countries~\cite{roediger_iii_competing_2019}.

Whereas oral tradition formed the basis for collective memory in early human history, today the media play a key role in determining what and who is remembered, and how~\cite{edgerton_television_2001,kligler-vilenchik_setting_2014,neiger_media_2011,zelizer_covering_1992}.
Researchers have studied the role of numerous media in constructing the post-mortem memory of deceased public figures.
A large body of work has investigated the journalistic format of the obituary~\cite{barry_epitaph_2008,fowler_obituary_2007,hume_obituaries_2000,starck_death_2008,hanusch_valuing_2008}, which captures how persons are remembered around the time of their death~\cite{hume_obituaries_2000}.
Taking a more long-term perspective, other work has considered how deceased public figures are remembered in the media over the course of years and decades~\cite{au_yeung_studying_2011,huet_mining_2013,cook_your_2012,van_de_rijt_only_2013,suchanek_semantic_2014}.
As ever more aspects of life are shifting to the online sphere, the Web is also gaining importance as a global memory place~\cite{pentzold_fixing_2009},
which has led researchers to study, \eg, how
social media users~\cite{gach_control_2017,stone_trauma_2002,sanderson_tweeting_2010,goh_analysis_2011,radford_grief_2012}
and Wikipedia editors~\cite{keegan_is_2015}
react to the death of public figures.
In the context of social media, the detailed analysis of highly visible individual cases, such as
Princess Diana~\cite{stone_trauma_2002},
pop star Michael Jackson~\cite{sanderson_tweeting_2010,goh_analysis_2011},
or race car driver Dale Earnhardt~\cite{radford_grief_2012},
has revealed how people experience and overcome the collective trauma that can ensue following the death of celebrities.

Although such studies of individuals have led to deep insights at a fine level of temporal granularity, they lack breadth by excluding all but some of the very most prominent public figures. What is largely absent from the literature is a general understanding of patterns of post-mortem memory in the media that goes beyond single public figures.

To bridge this gap, we draw inspiration from a body of related work that has studied the temporal evolution of collective memory using large-scale datasets---though, unlike our work, not with a focus on the immediate post-mortem period of public figures.
For instance, van de Rijt \etAl~\cite{van_de_rijt_only_2013} tracked thousands of person names in news articles, finding that famous people tend to be covered by the news persistently over decades.
In a similar analysis, Cook \etAl~\cite{cook_your_2012} further showed that the duration of fame had not decreased over the course of the last century.
Beyond news corpora, the online encyclopedia Wikipedia has become a prime resource for the data-driven study of collective memory.
Researchers have leveraged the textual content of Wikipedia articles~\cite{ferron_psychological_2012}, as well as logs of both editing~\cite{ferron_beyond_2014} and viewing~\cite{kanhabua_what_2014,garcia-gavilanes_memory_2017}, as proxies for the collective memory of traumatic events such as terrorist attacks or airplane crashes.
Jatowt \etAl~\cite{jatowt_digital_2016} characterized the coverage and popularity of historical figures in Wikipedia, observing vastly increased page-view counts for people from the 15th and 16th centuries,
a fact that Jara-Figueroa \etAl~\cite{jara-figueroa_how_2019} later attributed to the invention of the printing press.
In addition to news and encyclopedic articles,
books~\cite{michel_quantitative_2011,greenfield_changing_2013,pechenick_characterizing_2015}
and social media~\cite{jatowt_mapping_2015,sumikawa_digital_2018}
have also emerged as important assets for studying collective memory.

Whereas the above works are primarily descriptive in nature, researchers have also developed mathematical models of the growth and decay of collective memory.
Notably, as part of a rich literature on the evolution of performance, fame, and success in the arts and sciences~\cite{fortunato_science_2018,wang_quantifying_2013,fraiberger_quantifying_2018,sinatra_quantifying_2016,liu_hot_2018,yucesoy_untangling_2016},
Candia \etAl~\cite{candia_universal_2019} analyzed thousands of papers, patents, songs, movies, and athletes, showing that the decay of the intensity of collective memory can be well described by a biexponential function that captures two aspects of collective memory:
communicative memory, which is ``sustained by the oral transmission of information,'' and cultural memory, which is ``sustained by the physical recording of information''~\cite{candia_universal_2019}.

We extend this literature by studying how the coverage of thousands of public figures in news and social media evolved during the year following their death.
Our approach combines the Freebase knowledge base~\cite{bollacker_freebase_2008}---a comprehensive repository containing records for over 3~million public figures---with an extensive corpus of online news and social media compiled via the online media aggregation service Spinn3r~\cite{spinn3r_doc}, which comprises, for each day, hundreds of thousands of news articles from a complete set of all 6\,608 English-language Web domains indexed by Google News and tens of millions of social media posts from Twitter, amounting to about one third of full English Twitter (details in \textit{Materials and Methods;} number of documents per day in \SupFigNumDocs).
The population of study consists of \N public figures who died between 2009 and 2014 and received at least a minimum amount of pre-mortem coverage both in the news and on Twitter.
For each person, we tracked the daily frequency with which they were mentioned in the two media during the year before and the year after death,
and operationalize post-mortem memory via the resulting time series of mention frequency.
(For details about data and preprocessing, see \textit{Materials and Methods}.)
Analyzing the mention time series allowed us to quantify the extreme spike and rapid decay of attention that tend to follow the death of public figures, a pattern well captured by
\trackedChangeNew{a power law shifted by a constant additive offset.}
A cluster analysis of mention time series revealed four prototypical patterns of post-mortem memory (``blip'', ``silence'', ``rise'', and ``decline''),
and a regression analysis shed light on the biographic correlates of post-mortem memory and on systematic differences between post-mortem memory in mainstream news \vs\ social media.
We conclude that the prototypical persona with the largest post-mortem boost in English-language media attention can be described as an anglophone who was already well-known before death and died a young, unnatural death.
Long-term attention boosts
are on average smallest for leaders and largest for artists.
Finally, while both the mainstream news and Twitter are triggered by young and unnatural deaths, the mainstream news---but not Twitter---appears to also assume an additional role as stewards of collective memory when an old person or an accomplished leader dies.
Overall, the present work helps illuminate an age-old question: Who is remembered by society?

\section*{Results}

We strive to characterize the patterns by which post-mortem memory evolves during the year immediately following the death of public figures.
When considering this time frame, prior work has primarily taken a qualitative stance, asking \textit{how,} linguistically, the mainstream and social media speak about small sets of deceased people~\cite{starck_death_2008,hanusch_valuing_2008,gach_control_2017,stone_trauma_2002,sanderson_tweeting_2010,goh_analysis_2011,radford_grief_2012}.
In contrast, enabled by a comprehensive corpus of news and social media posts, we take a quantitative stance, asking \textit{about whom} the media speak \textit{how much} after death.

\begin{figure}
\centering
\subfigure[News]{
\label{fig:gallery_NEWS}
\includegraphics[width=0.25\linewidth]{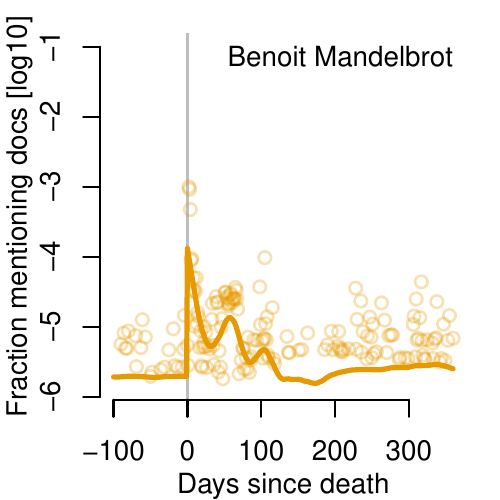}
\hspace{-4mm}
\includegraphics[width=0.25\linewidth]{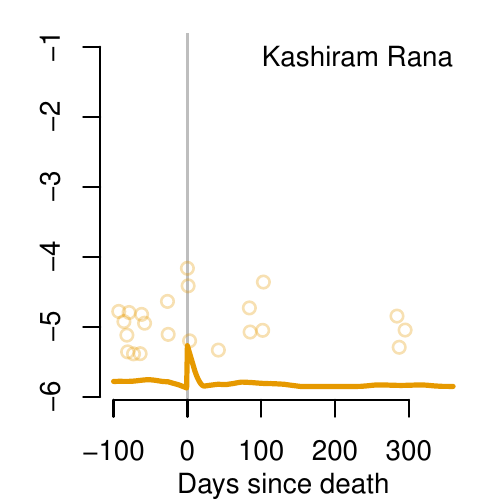}
\hspace{-4mm}
\includegraphics[width=0.25\linewidth]{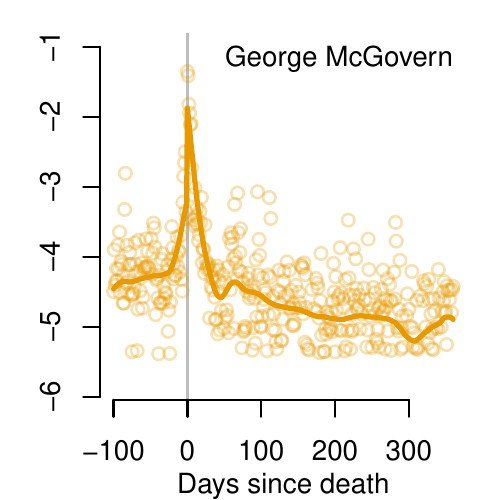}
\hspace{-4mm}
\includegraphics[width=0.25\linewidth]{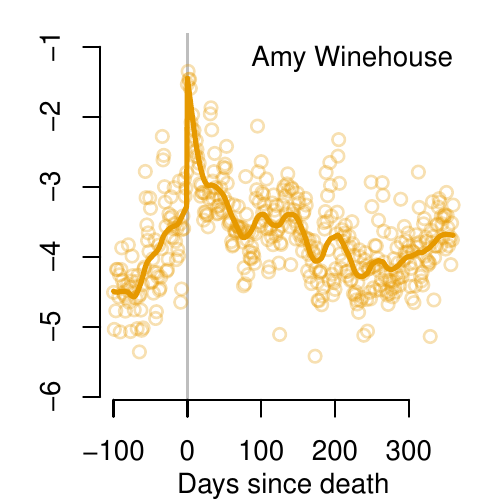}
}

\subfigure[Twitter]{
\label{fig:gallery_TWITTER}
\includegraphics[width=0.25\linewidth]{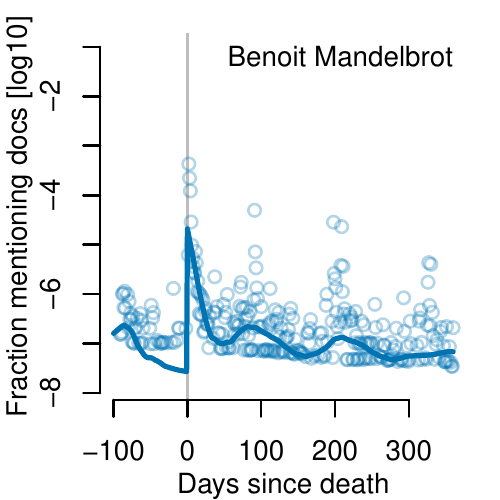}
\hspace{-4mm}
\includegraphics[width=0.25\linewidth]{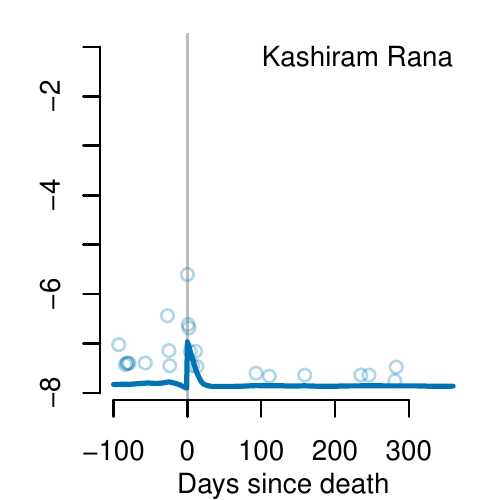}
\hspace{-4mm}
\includegraphics[width=0.25\linewidth]{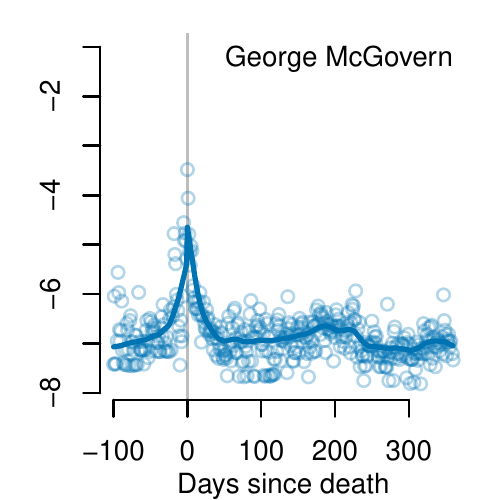}
\hspace{-4mm}
\includegraphics[width=0.25\linewidth]{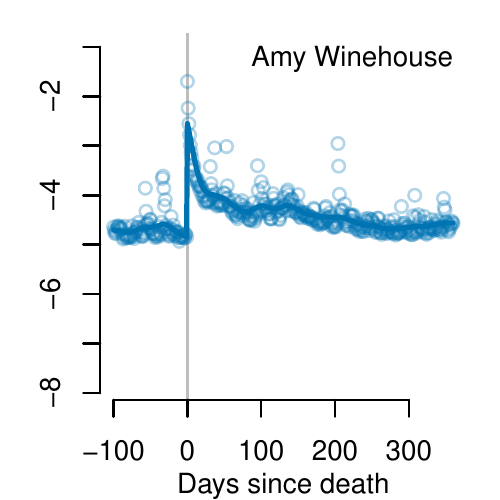}
}
\caption{
\textbf{Examples of mention time series for four deceased public figures,}
as observed \textbf{(a)}~in the news and \textbf{(b)}~on Twitter.
In mention time series, the $x$-axis specifies the number of days since death, and the $y$-axis, the base-10 logarithm of the fraction of documents in which the person was mentioned that day, out of all documents published that day.
Light circles correspond to raw mention time series, dark curves, to their smoothed versions.
}
\label{fig:gallery}
\end{figure}

At the core of our analysis are time series of mention frequency.
A person $i$'s \textit{raw mention time series} specifies, for each day $t$ relative to $i$'s day of death ($t=0$), the base-10 logarithm of the fraction
$S_i(t)$
of documents in which person $i$ was mentioned, out of all documents published on day~$t$.
To reduce noise, we also generated \textit{smoothed mention time series} using a variable span smoother based on local linear fits \cite{friedman_variable_1984}.
For each person, separate time series were computed for the news and for Twitter (examples in \Figref{fig:gallery}; additional details about mention time series construction in \textit{Materials and Methods}.)

\subsubsection*{The shape of post-mortem memory: communicative and cultural memory}

\begin{figure*}
\centering
\subfigure[News]{
\label{fig:avg_mention_curve_NEWS}
\includegraphics[width=0.47\textwidth]{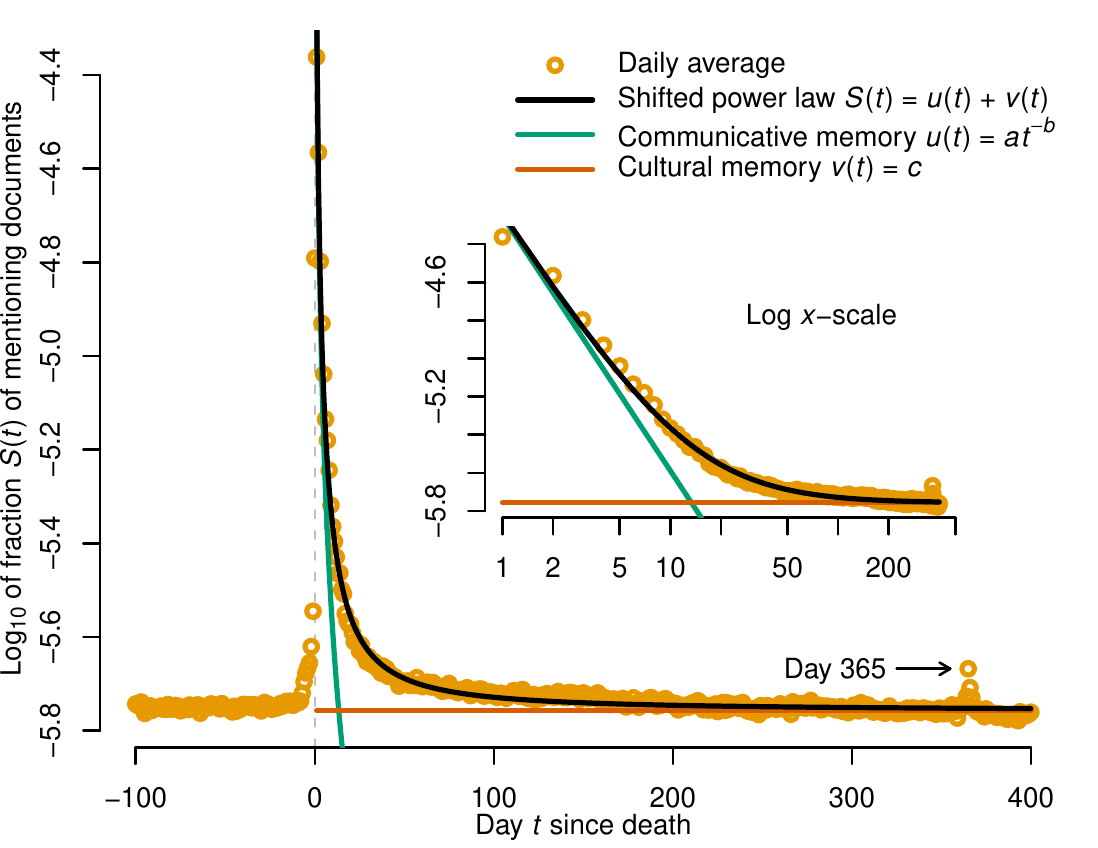}
}
\subfigure[Twitter]{
\label{fig:avg_mention_curve_TWITER}
\includegraphics[width=0.47\textwidth]{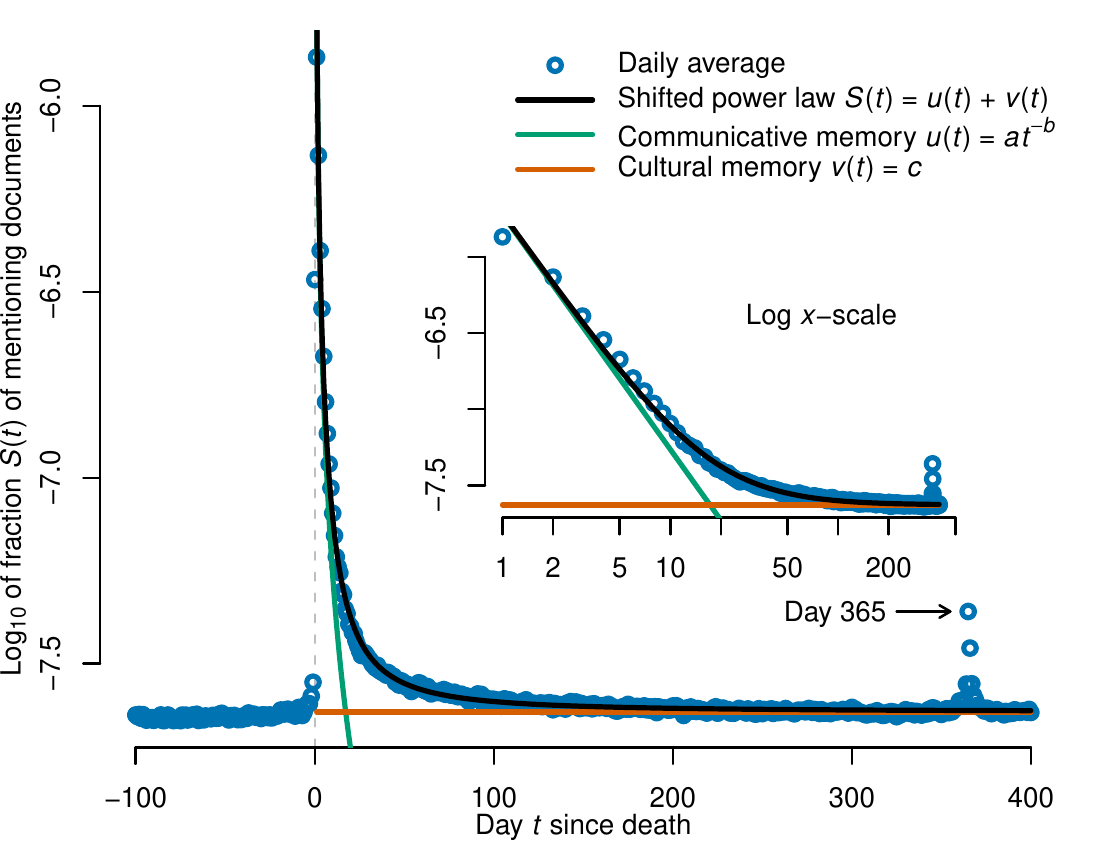}
}
\caption{
\textbf{Average mention time series,}
obtained via the arithmetic mean of the individual raw mention time series of the \N people included in the study,
\textbf{(a)}~in the news and
\textbf{(b)}~on Twitter
(see \Figref{fig:gallery} for examples of individual mention time series).
On average, the mention frequency of deceased public figures spikes by about 9\,400\% in the news, and by about 28\,000\% on Twitter, when they die, and fades quickly thereafter, with a minor secondary spike on the death anniversary.
\trackedChange{
We also plot the best fit of the shifted power law model (\textit{black}),
which decomposes the total collective memory $S(t)=u(t)+v(t)$ on day $t$
into a sum of
communicative memory $u(t)=at^{-b}$ (\textit{green})
and cultural memory $v(t)=c$ (\textit{red}).
}
Insets show the same data and fits on logarithmic $x$-axes.
}
\label{fig:avg_mention_curve}
\end{figure*}

Averaging the \N raw mention time series (\Figref{fig:avg_mention_curve}) exposes a sharp spike in the interest in public figures in the immediate wake of their death (on days 0 and 1),
followed by a steep drop up until around day 30,
where the curves elbow into a long, much flatter phase, which is only slightly disrupted by a small secondary spike on day 365 after death.
The main spike is so strong that, without logarithmically transforming the fractions $S_i(t)$ of mentioning documents, no interesting information besides the dominant main spike would be visually discernible.

\trackedChangeNew{
In a model that is conceptually similar to Candia \etAl's biexponential model \cite{candia_universal_2019}, we decompose the post-mortem collective memory $S(t)$ into a sum of two components, $S(t)=u(t)+v(t)$,
where $u(t)$ captures \textit{communicative memory,} and $v(t)$, \textit{cultural memory} (see introduction).
Communicative memory is modeled as a decaying power law, \ie, $u(t) = a t^{-b}$,%
\footnote{With $u(t)=a t^{-b}$, we have $\d u(t) / \d t = -(b/t)\ u(t)$, so the forgetting rate $b/t$ (fraction of memory lost at time $t$) decreases over time, in line with what is suggested by the psychological literature~\cite{wixted_form_1991}.}
whereas cultural memory is modeled as constant during the time frame considered here (400 days after death), \ie, $v(t)=c$.
We refer to this model as a \textit{shifted power law}.
It fits the empirical average mention time series ($R^2=0.99$; details on model fitting in \textit{Materials and Methods}) significantly better than any of eight alternative models from the literature~\cite{candia_universal_2019,rubin_one_1996,wixted_form_1991} (details in \SupFigAltModels{}), including the biexponential model \cite{candia_universal_2019}.
The best shifted power law fit is shown as a black line in \Figref{fig:avg_mention_curve}; the communicative and cultural memory components are plotted separately in green and red, respectively.

The fitted decay parameter of communicative memory is similar
for the news ($b=1.34$) and
for Twitter ($b=1.54$).
Communicative memory starts high on day $t=1$, but drops below cultural memory quickly, after 14 and 18 days in the news and on Twitter, respectively,
and accounts for only 25\%
of total collective memory after 31 and 36 days, respectively, which constitutes an inflection point where communicative memory levels off strongly (\SupFigCultVsCommMem).
Moreover, even though no pre-mortem data was used in fitting the model, the constant cultural memory $c$ closely approximates the average pre-mortem fraction of mentioning documents in both media (cf.\ \Figref{fig:avg_mention_curve}).

This suggests that, on average, public figures build up a certain baseline amount of (cultural) memory during their lifetime, on top of which a burst of quickly fading communicative memory is layered in the wake of death.
Note that, although collective memory rapidly reverts to the pre-mortem level when averaging over all people, it need not be so for individual people, as we shall see below (\Figref{fig:megafigure}(b)).}

Further evidence in support of two distinct memory modes comes from the fact that the average length of documents that mention a public figure dropped sharply with death (possibly due to brief death notes and obituaries) and reached the pre-mortem level again after about 30 days (\SupFigDocLength{}), \ie, around the inflection point where communicative memory levels off according to the fitted model.

We hence divide the post-mortem period into two phases: short-term (days 0 through 29) and long-term (days 30 through~360).
Based on this distinction, in order to reason about the shape of mention time series, we summarize each time series by four characteristic numbers (depicted graphically in \Figref{fig:megafigure}(a)):

\begin{figure*}
\centering
\includegraphics[width=\linewidth]{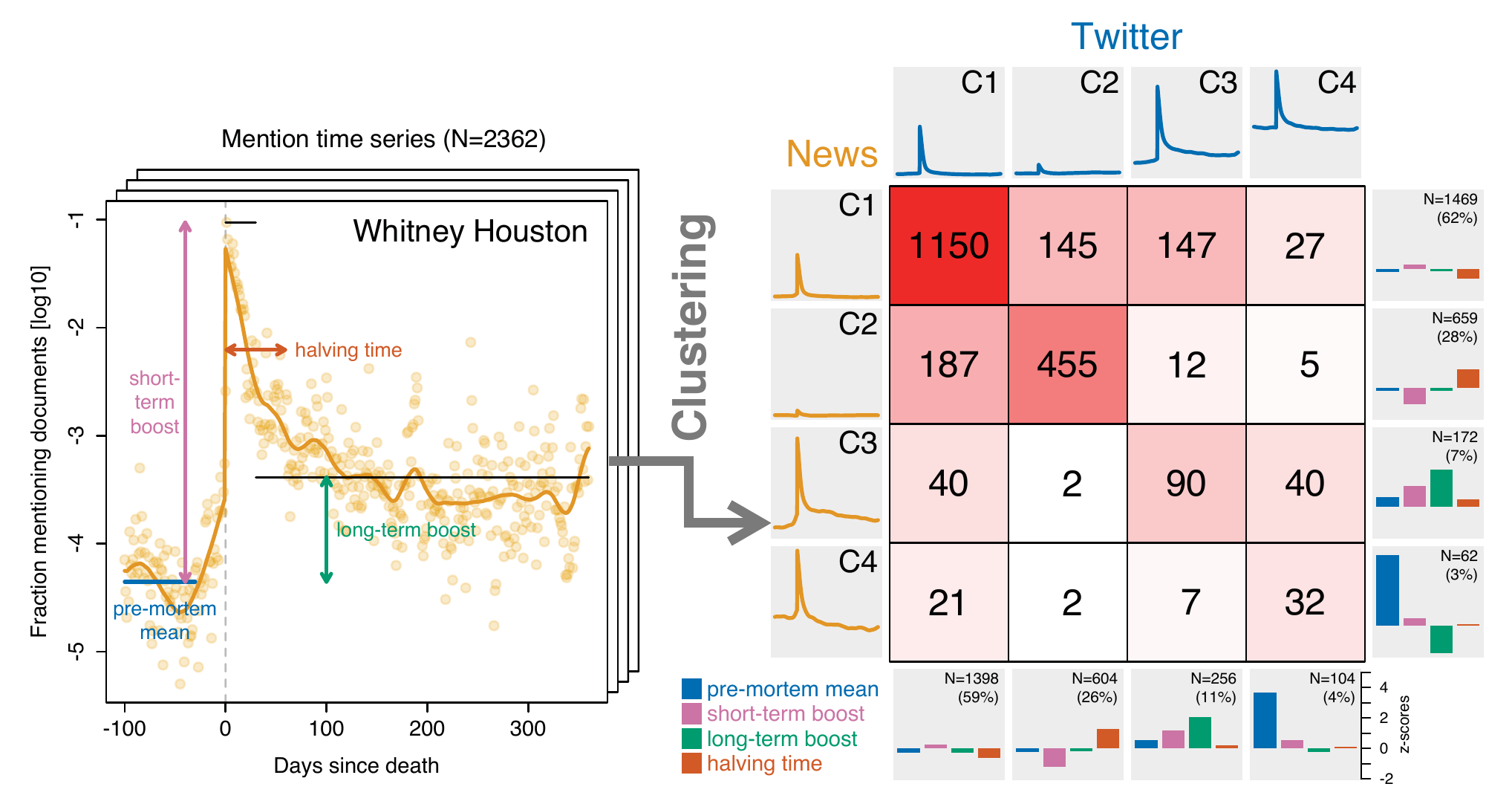}

(a) \hspace{85mm} (b)
\caption{
\textbf{Cluster analysis of mention time series.}
\textbf{(a)}~From each mention time series, we extract four characteristic numbers:
pre-mortem mean, short-term boost, long-term boost, and halving time.
In the resulting four\hyp dimensional space, time series are clustered using the $k$-means algorithm.
According to the average silhouette criterion, the optimal number of clusters is $k=4$ both in the news and on Twitter.
\textbf{(b)}~Nearly identical clusters (C1 through C4) emerge independently in the news and on Twitter, in terms of both cluster centroids and cluster sizes.
Cluster centroids are depicted as bar charts in the right (news) and bottom (Twitter) margins; average mention time series for each cluster, in the left (news) and top (Twitter) margins.
As captured by the confusion matrix, whose diagonal entries are much larger than under a null model that assumes the two media to be independent, a given person tends to fall into the corresponding clusters in the two media.
}
\label{fig:megafigure}
\end{figure*}

\begin{sidewaystable}
\begin{center}
\caption{
Linear regression models of short- and long-term boosts in news and Twitter. Standard errors of coefficients in parentheses.
}
\label{table:coefficients}

\begin{tabular}{l D{)}{)}{11)3} D{)}{)}{11)3} D{)}{)}{11)3} D{)}{)}{11)3} }
\toprule
 & \multicolumn{1}{c}{\shortstack{Short-term boost\\(News)}} & \multicolumn{1}{c}{\shortstack{Short-term boost\\(Twitter)}} & \multicolumn{1}{c}{\shortstack{Long-term boost\\(News)}} & \multicolumn{1}{c}{\shortstack{Long-term boost\\(Twitter)}} \\
\midrule
(Intercept)                           & 2.322 \; (.063)^{***} & 2.670 \; (.067)^{***} & .088 \; (.014)^{***}  & .095 \; (.015)^{***}  \\
Pre-mortem mean (relative rank)       & .804 \; (.093)^{***}  & .948 \; (.100)^{***}  & .031 \; (.020)        & .086 \; (.022)^{***}  \\
Manner of death: unnatural            & .618 \; (.095)^{***}  & .282 \; (.100)^{**}   & .097 \; (.021)^{***}  & .106 \; (.022)^{***}  \\
Language: non-anglophone              & -.316 \; (.074)^{***} & -.116 \; (.078)       & -.061 \; (.016)^{***} & -.037 \; (.017)^{*}   \\
Language: unknown                     & -.446 \; (.086)^{***} & -.325 \; (.091)^{***} & -.079 \; (.019)^{***} & -.081 \; (.020)^{***} \\
Gender: female                        & .083 \; (.072)        & -.034 \; (.076)       & .034 \; (.016)^{*}    & .006 \; (.017)        \\
Notability type: academia/engineering & .181 \; (.197)        & .340 \; (.208)        & -.032 \; (.043)       & .023 \; (.046)        \\
Notability type: general fame         & .070 \; (.124)        & .132 \; (.131)        & -.010 \; (.027)       & -.008 \; (.029)       \\
Notability type: known for death      & -.107 \; (.099)       & -.088 \; (.106)       & -.021 \; (.022)       & .008 \; (.023)        \\
Notability type: leadership           & .152 \; (.083)        & .113 \; (.087)        & -.058 \; (.018)^{**}  & -.040 \; (.019)^{*}   \\
Notability type: sports               & .049 \; (.083)        & .072 \; (.088)        & -.034 \; (.018)       & -.034 \; (.020)       \\
Age: 20--29                           & .162 \; (.170)        & .718 \; (.180)^{***}  & .060 \; (.037)        & .192 \; (.040)^{***}  \\
Age: 30--39                           & .400 \; (.167)^{*}    & .649 \; (.177)^{***}  & .028 \; (.037)        & .118 \; (.039)^{**}   \\
Age: 40--49                           & -.046 \; (.126)       & .351 \; (.133)^{**}   & -.001 \; (.028)       & .100 \; (.030)^{***}  \\
Age: 50--59                           & -.075 \; (.099)       & .181 \; (.104)        & -.058 \; (.022)^{**}  & -.024 \; (.023)       \\
Age: 60--69                           & -.109 \; (.082)       & .130 \; (.086)        & -.050 \; (.018)^{**}  & -.025 \; (.019)       \\
Age: 80--89                           & .022 \; (.078)        & .021 \; (.082)        & -.018 \; (.017)       & -.013 \; (.018)       \\
Age: 90--99                           & .174 \; (.098)        & .034 \; (.103)        & -.011 \; (.021)       & -.024 \; (.023)       \\
\midrule
R$^2$                                 & .213                  & .192                  & .123                  & .178                  \\
Adj. R$^2$                            & .197                  & .176                  & .106                  & .161                  \\
Num. obs.                             & 870                   & 870                   & 870                   & 870                   \\
RMSE                                  & .772                  & .815                  & .169                  & .181                  \\
\bottomrule
\multicolumn{5}{l}{\scriptsize{$^{***}p<0.001$, $^{**}p<0.01$, $^*p<0.05$}}
\end{tabular}

\end{center}
\end{sidewaystable}


\begin{table}
\begin{center}
\caption{
Linear regression models of news-minus-Twitter difference in short- and long-term boosts.
Standard errors of coefficients in parentheses.
}
\label{table:coefficients News vs. Twitter}

\begin{tabular}{l D{)}{)}{11)3} D{)}{)}{11)3} }
\toprule
 & \multicolumn{1}{c}{Short-term boost} & \multicolumn{1}{c}{Long-term boost} \\
\midrule
(Intercept)                           & -.427 \; (.047)^{***} & -.015 \; (.014)       \\
Pre-mortem mean (relative-rank diff.) & -.212 \; (.083)^{*}   & -.034 \; (.025)       \\
Manner of death: unnatural            & .348 \; (.070)^{***}  & -.008 \; (.021)       \\
Language: non-anglophone              & -.219 \; (.054)^{***} & -.023 \; (.017)       \\
Language: unknown                     & -.052 \; (.063)       & .012 \; (.019)        \\
Gender: female                        & .091 \; (.053)        & .029 \; (.016)        \\
Notability type: academia/engineering & -.048 \; (.146)       & -.046 \; (.044)       \\
Notability type: general fame         & -.016 \; (.092)       & .002 \; (.028)        \\
Notability type: known for death      & .105 \; (.074)        & -.015 \; (.022)       \\
Notability type: leadership           & .200 \; (.062)^{**}   & -.006 \; (.019)       \\
Notability type: sports               & .059 \; (.062)        & .009 \; (.019)        \\
Age: 20--29                           & -.577 \; (.126)^{***} & -.135 \; (.038)^{***} \\
Age: 30--39                           & -.235 \; (.124)       & -.089 \; (.038)^{*}   \\
Age: 40--49                           & -.374 \; (.093)^{***} & -.101 \; (.028)^{***} \\
Age: 50--59                           & -.204 \; (.073)^{**}  & -.029 \; (.022)       \\
Age: 60--69                           & -.175 \; (.061)^{**}  & -.021 \; (.018)       \\
Age: 80--89                           & .014 \; (.058)        & -.006 \; (.018)       \\
Age: 90--99                           & .164 \; (.072)^{*}    & .015 \; (.022)        \\
\midrule
R$^2$                                 & .101                  & .052                  \\
Adj. R$^2$                            & .083                  & .034                  \\
Num. obs.                             & 870                   & 870                   \\
RMSE                                  & .571                  & .174                  \\
\bottomrule
\multicolumn{3}{l}{\scriptsize{$^{***}p<0.001$, $^{**}p<0.01$, $^*p<0.05$}}
\end{tabular}

\end{center}
\end{table}

\begin{enumerate}
\item \textbf{Pre-mortem mean:} arithmetic mean of days 360 through 30 before death.
\item \textbf{Short-term boost:} maximum of days 0 through 29 after death, minus the pre-mortem mean.
\item \textbf{Long-term boost:} arithmetic mean of days 30 through 360 after death, minus the pre-mortem mean.
\item \textbf{Halving time:} number of days required to accumulate half of the total area between the post-mortem curve (including the day of death) and the minimum post-mortem value.
\end{enumerate}
All characteristics were computed on the smoothed time series, with the exception of the maximum used in the short-term boost, which was computed on the raw time series.
The 29 days immediately before death were excluded from the pre-mortem mean in an effort to exclude a potential rise in interest in people whose impending death might have been anticipated, \eg, due to illness.
Since the time series capture logarithmic mention frequencies, the (arithmetic) pre-mortem mean corresponds to the logarithm of the geometric mean mention frequency, and the short- and long-term boosts, to the logarithm of the multiplicative increase over the pre-mortem geometric mean mention frequency.

\subsubsection*{Magnitude of short- and long-term boosts}

Aggregating the short-term boost over all public figures allows us to quantify the sharp spike immediately after death observed in \Figref{fig:avg_mention_curve}.
The median short-term boost was 1.98 (95\% CI $[1.90,2.03]$) in the news,
and 2.45 (95\% CI $[2.37,2.50]$) on Twitter.
(All curve characteristics are summarized in \SupTabCurveChars{} and \SupFigCurveChars.)
The boost was significantly stronger on Twitter
(Wilcoxon's signed-rank test: $W=477\,893$, two-sided $p < 10^{-15}$),
where it approximately corresponded to a 28\,000\% increase on the linear scale
($10^{2.45} \approx 281$),
compared to a 9\,400\% increase in the news
($10^{1.98} \approx 95$).

After the immediate spike, media interest tended to fade quickly.
In the news, no important long-term boost was observed (median 0.000545, 95\% CI $[-0.000908, 0.00171]$),
whereas on Twitter, we measured a significantly larger
(Wilcoxon's signed-rank test: $W=881\,590$, two-sided $p < 10^{-15}$)
long-term boost of 0.0160 in the median (95\% CI $[0.0133, 0.0175]$),
translating to a 3.8\% increase on the linear scale
($10^{0.016} \approx 1.038$).

\subsubsection*{Cluster analysis of mention time series}

Mention time series expose a great variety of curve shapes, a glimpse of which is given by the examples of \Figref{fig:gallery}.
We hypothesized that, despite their diversity, mention time series could be grouped into distinct classes, a hypothesis that we explored in a cluster analysis.
Time series were represented by their four characteristic numbers
(pre-mortem mean,
short-term boost,
long-term boost,
halving time)
in $z$-score\hyp standardized form and clustered using the $k$-means algorithm.
A separate clustering was performed for the news and for Twitter.
Evaluating all numbers of clusters $k \in \{2, \dots, 30\}$ via the average silhouette criterion \cite{rousseeuw_silhouettes_1987} revealed a clear optimum for $k=4$ clusters for both the news and Twitter (\SupFigSilhouette).

The cluster centroids are visualized in the right and bottom margins of \Figref{fig:megafigure}(b);
the right margin shows the centroids for the news,
the bottom margin, for Twitter.
Moreover, we plot the average smoothed mention time series for each cluster in the left (news) and top (Twitter) margins.
(An overlay of all time series per cluster is plotted in \SupFigClusterOverlay.)
Strikingly, although the clustering was performed independently for the news and for Twitter, respectively, the centroids that emerged---as well as the number of data points in each cluster---are nearly identical.
The resulting clusters, which we name C1 through C4 in order of decreasing size, can be described as follows:

\begin{enumerate}
\item[\textbf{C1}] \textbf{(``blip''):} Average mention frequency pre- as well as post-mortem, with a short-term boost of average magnitude in between (62\% of people in the news; 59\% on Twitter).
\item[\textbf{C2}] \textbf{(``silence''):} Average mention frequency pre- as well as post-mortem, with a faint short-term boost of below-average magnitude in between (28\% in the news; 26\% on Twitter).
\item[\textbf{C3}] \textbf{(``rise''):} High pre-mortem mention frequency, large short-term boost, followed by an extreme long-term boost (7\% in the news; 11\% on Twitter).
\item[\textbf{C4}] \textbf{(``decline''):} Extreme pre-mortem mention frequency, above-average short-term boost, followed by a below-average long-term boost (3\% in the news; 4\% on Twitter).
\end{enumerate}

In both media, over half of the people (59--62\%) fall into cluster C1; their time series resemble the overall average (see \Figref{fig:avg_mention_curve}), with a brief spike after death and a quick drop to the---usually low---pre-mortem level.
About half of the remaining people (26--28\%) fall into cluster C2; their time series are similar to those of C1, with the difference that the death of people in C2 went largely unnoticed.
About half of the people outside of C1 and C2 (7--11\%) fall into C3, which mostly contains people who were popular already before death and experienced a large boost in attention in both the short and the long term.
The final cluster, C4, is composed of a tiny elite (3--4\%) of people of an extreme pre-mortem popularity that tended to fade post-mortem. The long-term decrease was considerably stronger in the news than on Twitter in this cluster.

Not only do nearly identical clusters of nearly identical size emerge in the news as on Twitter;
a given person also tends to fall into the corresponding clusters in the two media,
as captured by the cluster confusion matrix (\Figref{fig:megafigure}(b)), which counts, for all $i,j \in \{1,2,3,4\}$, the number of people falling into news cluster $i$ and Twitter cluster $j$.
Using Pearson's $\chi^2$ test, we reject the null hypothesis under which cluster membership is assumed to be independent in the news \vs\ Twitter, given the empirical marginal cluster sizes
($\chi^2 = 1\,739, p < 10^{-5}$).
%
%
In particular, all diagonal entries of the confusion matrix are strongly over\hyp represented,%
\footnote{
The empirical trace (the sum of diagonal elements) is 1\,727.
Under the null hypothesis, the expected trace would be 1\,059.
Given the marginal constraints, the minimum and maximum traces that could possibly be attained are 505 and 2\,236, respectively (determined via linear programming).
That is, the empirical trace attains 71\% of the range between the minimum and the maximum, whereas the null model attains only 32\%.
}
whereas all but two off-diagonal entries are under\hyp represented, and ``C3 in news, C4 on Twitter'' is the only off-diagonal entry to be significantly over\hyp represented
(one-sample proportions test with continuity correction: $\chi^2 = 135, p < 10^{-15}$).

\subsubsection*{Biographic correlates of post-mortem memory}

Next, we aim to understand what pre-mortem properties of a person are associated with their post-mortem mention frequency.
A na\"ive correlational analysis would not suffice for this purpose, as personal properties are correlated with one another; \eg, leaders (politicians, CEOs, etc.)\ in the dataset are more likely to have died old and of a natural death, and are more likely to be men, compared to artists.
In order to disentangle such correlations, we performed a regression analysis.
We fitted linear regression models for two outcomes:
\begin{enumerate}
\denselist
\item short-term boost,
\item long-term boost,
\end{enumerate}
and with six predictors in either case:
\begin{enumerate}
\denselist
\item pre-mortem mean mention frequency,
\item age at death (factor with eight levels: 20--29, 30--39, \dots, 90--99),
\item manner of death (factor with two levels: natural, unnatural),
\item notability type (factor with six levels, specifying a profession or role for which the person was most known:
arts,
sports,
\trackedChange{leadership [including politicians, business\slash organization leaders, religious leaders, military, etc.],}
\trackedChange{known for death [including disaster victims],}
general fame,
academia\slash engineering),
\item \trackedChange{language} (factor with three levels: anglophone, non-anglophone, unknown),
\item gender (factor with two levels: female, male).
\end{enumerate}

\begin{figure*}
\centering
\includegraphics[width=0.32\linewidth]{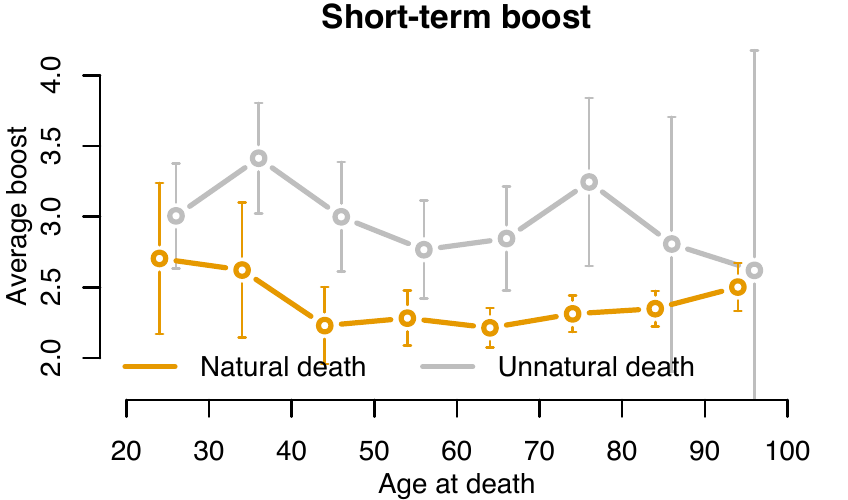}
\includegraphics[width=0.32\linewidth]{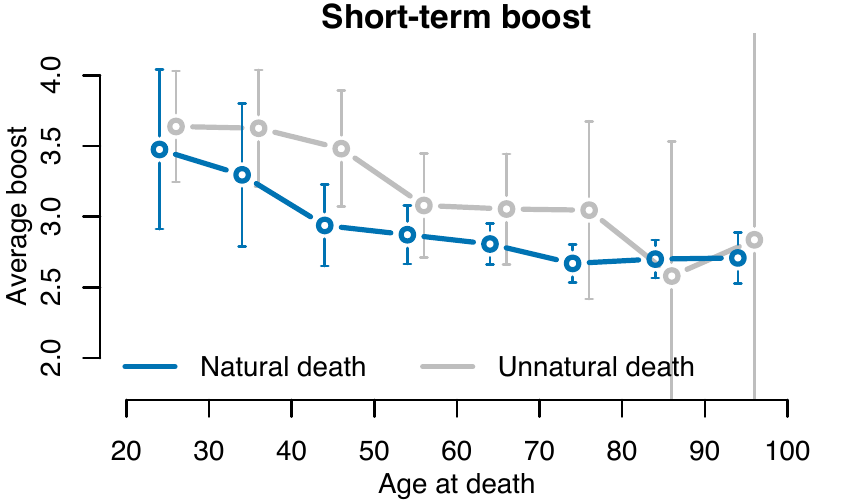}
\includegraphics[width=0.32\linewidth]{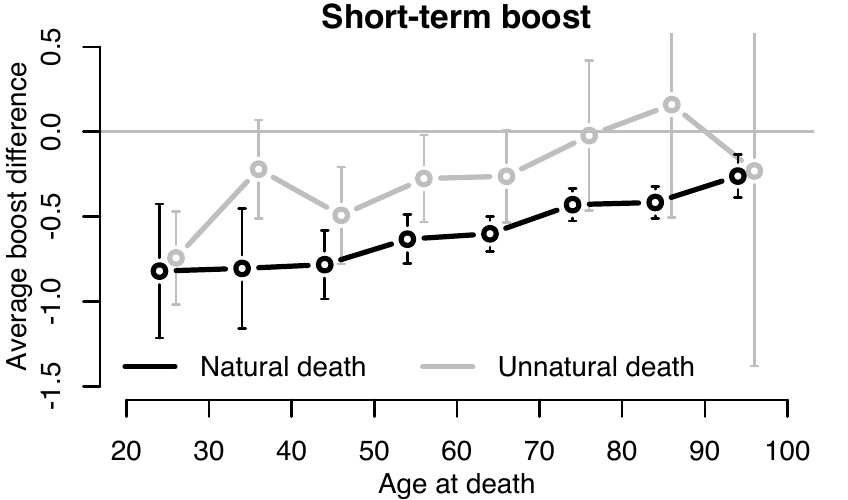}

\subfigure[News]{
\label{fig:age_coefs_NEWS}
\includegraphics[width=0.32\linewidth]{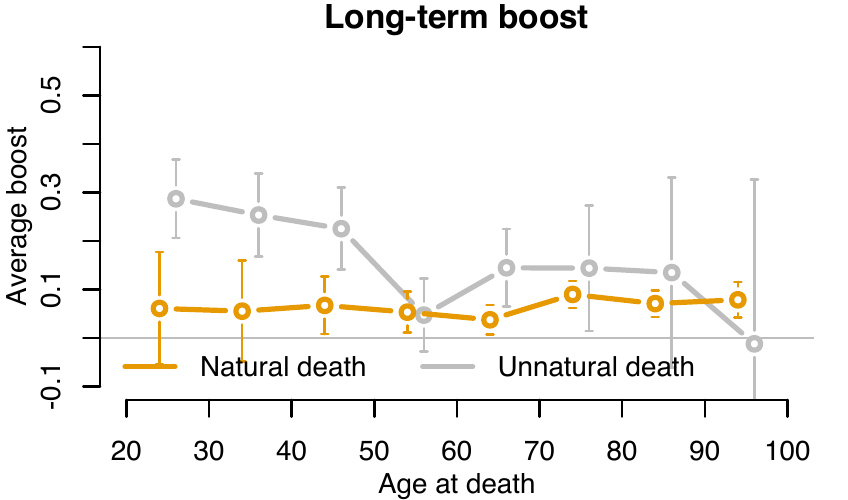}
}
\hspace{-3.3mm}
\subfigure[Twitter]{
\label{fig:age_coefs_TWITTER}
\includegraphics[width=0.32\linewidth]{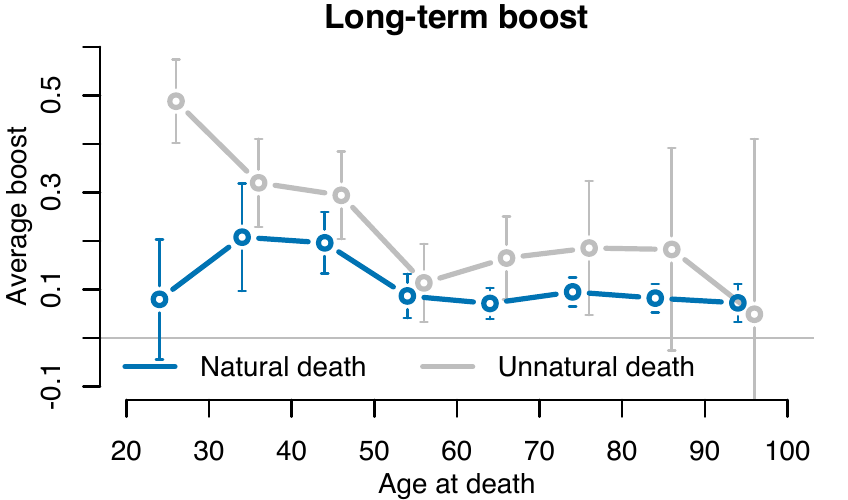}
}
\hspace{-3.3mm}
\subfigure[News \vs\ Twitter]{
\label{fig:age_coefs_NEWS_VS_TWITTER}
\includegraphics[width=0.32\linewidth]{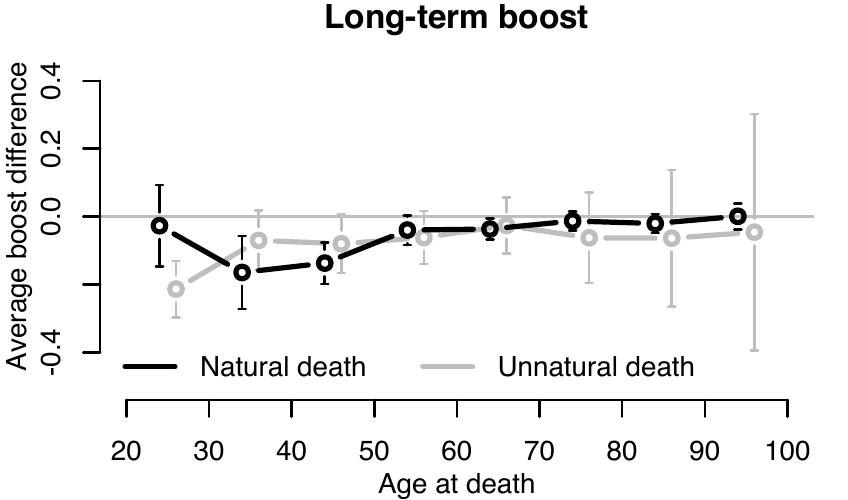}
}
\caption{
\textbf{Dependence of post-mortem mention frequency on age at death}
for
\textbf{(a)}~the news and
\textbf{(b)}~Twitter,
in terms of short-term (\textit{top}) and long-term (\textit{bottom}) boost,
\trackedChange{defined as the base-10 logarithm of the post-to-pre-mortem ratio of fractions of mentioning documents.}
Results were obtained via linear regression models that controlled for pre-mortem mean, notability type, \trackedChange{language}, and gender.
\textbf{(c)}~Per-person news-minus-Twitter difference in short-term (\textit{top}) and long-term (\textit{bottom}) boosts.
Error bars capture 95\% confidence intervals (approximated as $\pm$2 standard errors).
These plots show
that unnatural deaths lead to larger attention boosts across age brackets, both short- and long-term;
that the news increases attention most for those who die very young or very old, whereas Twitter increases attention more the younger the deceased person;
and that the difference between attention boosts in news \vs\ Twitter is larger for those who die older (short- and long-term) and for those who die an unnatural death, across age brackets (short-term).
}
\label{fig:age_coefs}
\end{figure*}

Out of all \N people, the regression analysis only included the 870 people for whom all factors took one of the above\hyp defined levels (details on definition of biographic features in \textit{Materials and Methods;} distribution summarized in in \SupTabBioStats{}).
All factor variables were dummy-coded as binary indicators.
The pre-mortem mean was first rank-transformed and then linearly scaled and shifted to the interval $[-0.5,0.5]$.
Additionally, 70--79 years (which contains the mean and median age at death) was chosen as the default age level,
and the most frequent level was chosen as the default for all other factors, such that the regression intercept captures the average outcome for a ``baseline persona'' representing
male anglophone artists of median pre-mortem popularity who died a natural death at age 70--79.
\trackedChange{With the above, a coefficient $\beta$ for a binary predictor corresponds to an additive boost increase of $\beta$ with respect to the baseline persona,
or---since boosts are base-10 logarithms (of post-to-pre-mortem ratios of mention frequencies)---to a multiplicative post-to-pre-mortem ratio increase of $10^\beta$.}
A separate regression model was fitted for each combination of medium (news or Twitter) and outcome (short- or long-term boost), for a total of four models.

The model coefficients (summarized in \Tabref{table:coefficients}) paint a largely consistent picture for the news \vs\ Twitter.
We observe that, in both media, \textit{ceteris paribus,}
both the short- and the long-term boost was larger
for people who died an unnatural death,
for people with an anglophone background,
and for people who were already popular pre-mortem.
No significant gender variation was detected,
with the exception of the long-term boost in the news, which was slightly larger for women.
The only significant notability type was leadership,
\trackedChange{whose long-term boost was smaller than that of the baseline (arts) in both media.}

The dependence of short- and long-term boosts on the age at death is displayed visually in \Figref{fig:age_coefs_NEWS} and \ref{fig:age_coefs_TWITTER}.
In order to determine whether the above finding that attention increased more for people who died an unnatural death holds across age brackets, the plots are based on a slightly modified model with an additional ``age by manner of death'' interaction term.
This allows us to estimate the average post-mortem attention boost attained by each age bracket separately for people who died natural \vs\ unnatural deaths (as before, the estimates are for male anglophone artists of median pre-mortem popularity).
Inspecting the curves of \Figref{fig:age_coefs_NEWS} and \ref{fig:age_coefs_TWITTER}, we make two observations.
First, across age brackets, people who died an unnatural death received larger boosts, both short- and long-term, and both in the news and on Twitter.
Second, the curves have a non\hyp monotonic U-shape for the news (\Figref{fig:age_coefs_NEWS}), but are monotonically decreasing for Twitter (\Figref{fig:age_coefs_TWITTER});
\ie, the news increased attention most for those who died either very young or very old, whereas Twitter increased attention more the younger the deceased.

\subsubsection*{News \vs\ Twitter}

The above analyses were done separately for the news and Twitter.
In order to understand how post-mortem memory of the same person differed between the two media, we conducted a pairwise analysis.
We again fitted linear regression models with the same predictors as above, but this time with outcomes defined by the news-minus-Twitter difference in short- and long-term boosts.
Accordingly, the rank-transformed and scaled pre-mortem mean predictor was replaced with the news-minus-Twitter difference in rank-transformed and scaled pre-mortem means.
Given this setup, large positive coefficients mark groups of people who received particularly strong boosts in the news compared to Twitter, and large negative coefficients mark groups of people who received particularly strong boosts on Twitter compared to the news.

The model coefficients (summarized in \Tabref{table:coefficients News vs. Twitter}) reveal that those who died an unnatural death, as well as leaders, received
particularly large short-term attention boosts in the news compared to Twitter.
Conversely, pre-mortem popular people and those with a non\hyp anglophone background received
particularly large short-term attention boosts on Twitter compared to the news,
the latter possibly because English is the most globally connected language~\cite{ronen_links_2014}, such that Twitter posts, even though all written in English, stemmed from a more geographically and culturally diverse set of writers than news articles, which originated exclusively from anglophone outlets by design.
Other than leaders, no further notability type was significantly associated with either outcome, and no significant gender variation was observed.
Finally, the age dependence is visualized in \Figref{fig:age_coefs_NEWS_VS_TWITTER}, which shows that, the older a person, the larger the news-minus-Twitter difference in boosts, confirming that news media favored older people more than Twitter did, both short- and long-term.

\section*{Discussion}

Our analysis of mention frequencies over time revealed that, for the majority of public figures, a sharp pulse of media attention immediately followed death, whereby mention frequency increased by 9\,400\% in the news, and by 28\,000\% on Twitter, in the median.
The average mention frequency then declined sharply, with an inflection point around one month after death, from where on it decayed more slowly, eventually converging toward the pre-mortem level.
These two stages are consistent with a model that posits two components of collective memory:
\trackedChangeNew{
a constant baseline level of cultural memory built up during life,
and an added layer of communicative memory that is sparked by death and usually decays in a matter of days according to a power law.}
A cluster analysis of the mention time series revealed a set of four prototypical memory patterns (``blip'', ``silence'', ``rise'', and ``decline'').
The same set of patterns emerged independently in the news and on Twitter, and the same person tended to fall into the same cluster across the two media.

In our regression analysis of biographic correlates of post-mortem memory, out of all notability types (arts, sports, leadership, known for death, general fame, academia\slash engineering), only leadership (politicians, business leaders, \etc)\ stood out significantly, being associated with a particularly low boost in long-term memory (\Tabref{table:coefficients}).
One might wonder if this fact could simply be explained by a regression to the mean, since leaders had the highest pre-mortem mention frequencies in the news (\SupFigCurveCharsByType(a)). Note, however, that on Twitter, too, leaders saw the lowest long-term boosts,
despite the fact that, on Twitter, it is artists---not leaders---who had the highest pre-mortem mention frequencies (\SupFigCurveCharsByType(b)). We thus consider a regression to the mean an unlikely explanation for the low long-term boosts of leaders, as it could not simultaneously explain the situation in both media.
Rather, we speculate that the lives and legacies of people of different notability types might differ systematically.
Considering that nearly all (eight out of 10) long-term boost coefficients for notability types are negative in \Tabref{table:coefficients} (and none are significantly positive), the distinction to be made is in fact not that between leadership and the rest, but rather that between arts (the default notability type) and the rest.
Based on this observation, we hypothesize that artists remain more present in the collective memory because they not only tend to be active performers during their lifetime, but also frequently leave a legacy of artwork that can long survive them,
whereas leaders, athletes, \etc, are noteworthy primarily for the actions they take during their lifetime, and are of much decreased interest to the media once they cannot take action anymore---an effect that seems to be most pronounced for leaders.%
\footnote{In terms of interesting exceptions to this rule, certain leaders saw large long-term boosts in the news (\eg, Zimbabwean military Solomon Mujuru, who died in a fire in circumstances considered suspicious by some), on Twitter (\eg, World War II veterans Richard Winters and Maurice Brown), or in both media (\eg, entrepreneur and activist Aaron Swartz, suicide at age~26; Governor of Punjab Salman Taseer, assassinated at age~66).}


The low coefficients of determination (adjusted $R^2$) of the linear regression models, ranging from
0.106 to 0.197
(\Tabref{table:coefficients}),
serve as a testimony of the richness of human lives and legacies, which cannot be captured by statistical models relying on just a few biographic variables.
Given the inherent unpredictability of social systems~\cite{martin_exploring_2016,salganik_experimental_2006}, this would be unlikely to change even if more biographic variables and more data points became available, and if more complex statistical models were to be used.
We emphasize, however, that despite the inherent limits of predictability all model fits were highly significant
($p<10^{-15}$ for the $F$-statistics of all models of \Tabref{table:coefficients}, cf.\ \textit{Regression modeling} in Supplementary Information).
Also, and most importantly, the effects were not only significant, but also large.
\trackedChange{
As mentioned, a coefficient $\beta$ for a binary predictor corresponds to a multiplicative increase of $10^\beta$ in the post-to-pre-mortem ratio of mention frequencies, compared to the baseline persona, a male anglophone artist of median pre-mortem popularity who died a natural death at age 70--79.
For example, \textit{ceteris paribus,} an unnatural death quadrupled ($10^{0.618} \approx 4.15$) the short-term post-to-pre-mortem mention\hyp frequency ratio in the news, and nearly doubled it ($10^{0.282} \approx 1.92$) on Twitter.
The effect of age at death was also large.
For instance, on Twitter, \textit{ceteris paribus,} the short-term post-to-pre-mortem mention\hyp frequency ratio for the 30--39 age bracket was twice that of the neighboring, 40--49 age bracket ($10^{0.649-0.351} \approx 1.99$);
and that of the youngest age bracket was nearly five times that of the oldest age bracket ($10^{0.718-0.034} \approx 4.83$).
}


One of the key contributions of this study is the comparison between mainstream news and Twitter---a prominent social media platform---on a fixed set of attention subjects, thus extending a rich literature on the interplay between the two media \cite{grinberg_fake_2019,bovet_influence_2019,zhao_comparing_2011,kwak_what_2010}.
Despite the striking similarity of prototypical mention time series emerging from the cluster analysis (\Figref{fig:megafigure}), the regression analysis revealed several noteworthy differences between post-mortem memory in the news \vs\ Twitter.
First, whereas on Twitter the post-mortem boost was monotonically and negatively associated with age at death (\Figref{fig:age_coefs_TWITTER}),
we observed a non\hyp monotonic U-shaped relationship in the news (\Figref{fig:age_coefs_NEWS}), which provided the largest post-mortem boost to both those who died very young and to those who died very old, an effect that even held for a fixed person (\Figref{fig:age_coefs_NEWS_VS_TWITTER}).
Second, the increased short-term boost associated with unnatural deaths was even more pronounced in the news than on Twitter (\Tabref{table:coefficients News vs. Twitter}), across age groups (\Figref{fig:age_coefs_NEWS_VS_TWITTER}).
And third, leaders were boosted more by the news than by Twitter, both short- and long-term (\Tabref{table:coefficients News vs. Twitter}).
Taken together, these findings could be interpreted as the result of two simultaneous roles played by mainstream news media:
on the one hand, as heralds catering to the public curiosity stirred by a young or unnatural death;
on the other hand, as stewards of collective memory when an old person or an accomplished leader dies after a life of achievement.
On the contrary, the extent to which Twitter plays both roles is weaker:
on the one hand, unnatural deaths were followed by a less pronounced short-term boost on Twitter than in the news;
on the other hand, Twitter users paid less attention when an old public figure or a leader died.

The present study showed that even the simple counting of mentions yields nuanced insights into \textit{who} is remembered after death.
Future studies may go further by also asking \textit{how} deceased public figures are remembered, by studying how the language, tone, and attitude toward them change in the wake of death.
By considering a diverse set of thousands of public figures such as ours, future work will be able to quantify, \eg, to what extent news and social media abide by the old Latin adage \textit{``De mortuis nihil nisi bonum''} (``Of the dead, speak no evil'').
\trackedChange{The analysis could be further enriched by going beyond the coarse biographic categories considered here and leveraging manually curated repositories of more fine-grained information about public figures~\cite{yu_pantheon_2016}.}
\trackedChange{
We also emphasize that media attention cannot capture all aspects of collective memory, so we encourage researchers to apply our methodology to further measures of popularity, in particular those capturing the consumption, rather than production, of content, including songs, movies, books, Wikipedia articles, \etc
}

Finally, this study started from an elite of people considered worthy of being included in the Freebase knowledge base (which roughly equals the set of people with a Wikipedia article).
This notability bias was further increased by discarding people whose pre-mortem mention frequency was too low in the news or on Twitter (see \textit{Materials and Methods}), a restriction necessary in order to compare the coverage of a fixed person across the two media.
Since the bar for being mentioned in the news~\cite{shor_large-scale_2019} as well as for being included in Freebase and Wikipedia~\cite{wagner_women_2016} is higher for women than for men, the women included in the study are likely to be more inherently\hyp noteworthy than the men included.
This might in turn affect our estimate of the association of gender with post-mortem memory:
although we identified only small and mostly insignificant effects, it is conceivable that different effects might appear if the inherent noteworthiness was equalized across genders in the dataset by lowering the bar for inclusion for women or raising it for men---an important methodological challenge.

Going forward, researchers should also strive to remove the bar for inclusion in a study of post-mortem memory altogether, by moving from a noteworthy elite of public figures to a representative set of regular people.
With the widespread adoption of social media, we may, for the first time in history, not only ask, but also answer, who is remembered after they die.

\section*{Methods}
\subsubsection*{News and Twitter corpora}

We compiled a large corpus of media coverage via the online media aggregation service Spinn3r, which provides ``social media, weblogs, news, video, and live web content''~\cite{spinn3r_doc}.
We had full access to the Spinn3r data stream and collected a complete copy over the course of more than five years (June 2009 to September 2014) via the Spinn3r API, for a total of around 40~terabytes of data.
Besides the main text content, documents consist of a title, a URL, and a publication date.


Twitter posts (\textit{tweets}) were easy to recognize automatically in the Spinn3r data, whereas news articles were not explicitly labeled as such.
In order to identify news articles, we started from a comprehensive list of all 151K online news articles about Osama bin Laden's killing (2~May 2011) indexed by Google News \cite{bharat2011google}.
Assuming that every relevant news outlet had reported on bin Laden's death, we labeled as news articles all documents in the Spinn3r crawl that were published on one of the 6\,608 Web domains that also published an English news article about bin Laden's death according to the Google News list.

We included in our analysis all English-language news articles and tweets collected between 11~June 2009 and 30~September 2014. The resulting corpus comprises, for each day, hundreds of thousands of news articles and tens of millions of tweets (\SupFigNumDocs).

%
%

Although Spinn3r does not publicly disclose its data collection strategy, we assess the corpus as highly comprehensive (\SupTabTwitterCompleteness{}).

\subsubsection*{Detecting people mentions}

In order to construct mention time series (\Figref{fig:gallery}), we had to identify documents that contain the names of dead public figures.
This is not a trivial task, since names may be ambiguous. Entity disambiguation is a well-studied task, but unfortunately natural language processing--based methods were too resource\hyp intensive for our 40-terabyte corpus, so we resorted to a simpler method:
in addition to fully\hyp unambiguous names and aliases (henceforth simply ``names'') belonging to a single entity,
we identified a set of mostly\hyp unambiguous names, which refer to the same entity at least 90\% of the time in English Wikipedia, and we mapped each mention of such a name in the Spinn3r corpus to the entity it most frequently referred to in Wikipedia (people without any highly unambiguous name were excluded).
For Twitter, we considered a tweet to mention a given person if the tweet mentioned at least one full name of the person.
For the news, we additionally required at least one additional mention of the person (via a full name or a token-based prefix or suffix of a full name), in order to reduce spurious mentions (\eg, when the person was merely mentioned in a link to another article, rather than in the core article content).

\subsubsection*{Inclusion criteria}

In order to compile a set of dead public figures, we started from the 33\,340 people listed in the 28~September 2014 version of the Freebase knowledge base as having died during the period spanned by our media corpus (11~June 2009 to 30~September 2014).
On 86 out of these 1\,936 days, Spinn3r failed to provide data due to technical problems.
We excluded people for whom the 100 days immediately following death included at least one of the 86 missing days (in order to obtain better estimates of short- and long-term boosts), who died within less than 360 days of the corpus boundary dates (in order to compute pre-mortem means and long-term boosts in the same way for everyone), who were mentioned on fewer than five days in either of the news or Twitter during the 360 days before death (in order to avoid extremely noisy pre-mortem means), or whose names on English Wikipedia contained parentheses, \eg, ``John Spence (Trinidad politician)'' (as such names are unlikely to be used in prose).
These criteria reduced the set of people from 33\,340 to \N.


\subsubsection*{Biographic features}

Each public figure was described by the following biographic features, extracted or computed from Freebase:
age at death,
gender,
manner of death (natural or unnatural, inferred from the more detailed cause of death, cf.\ \textit{Taxonomy of causes of death} in Supplementary Information),
language (``anglophone'' for citizens of the U.S., Canada, the U.K., Ireland, Australia, New Zealand, or South Africa; ``non-anglophone'' for citizens of other countries; ``unknown'' for people with no nationality listed in Freebase),
and notability type (a profession or role for which the person was most known, \eg, ``singer'' for Whitney Houston; manually grouped into six classes:
arts, sports, leadership, known for death, general fame, academia\slash engineering;
cf.\ \textit{Taxonomy of notability types} in Supplementary Information).
The distribution of these features is summarized in \SupTabBioStats{} for all public figures, for the people included in the study, and for the subset retained for the regression analysis.

\subsubsection*{Mention time series}

To avoid taking logarithms of zero when constructing mention time series, a constant value of $\epsilon$ was added to each individual time series before taking logarithms, where $\epsilon$ was the minimum non-zero value across all individual time series (but note that the raw time series of \Figref{fig:gallery} were drawn without adding $\epsilon$).
Separate values of $\epsilon$ were computed for the news and for Twitter.
When smoothing time series via the variable span smoother~\cite{friedman_variable_1984}, we considered the pre- and post-mortem periods separately, in order to not smooth out the spike that usually immediately followed death.
Missing days were interpolated linearly.


\trackedChangeNew{
\subsubsection*{Model fitting}

In order to fit the shifted power law model to the data, we operate on the logarithmic scale, by finding the nonlinear least-squares estimates
\begin{equation}
\argmin_{a,b,c} \,
\sum_{t=1}^{400}
\left(
\langle \log S_i(t) \rangle -
\log \left( a t^{-b} + c  \right)
\right)^2,
\end{equation}
where $\langle \log S_i(t) \rangle$ is the arithmetic mean of the empirically measured $\log S_i(t)$ over all persons $i$.
The following optimal parameters were obtained:
\begin{eqnarray}
\text{News:} & a=5.58 \times 10^{-5}, \;\;\; b=1.34, \;\;\; c=1.75 \times 10^{-6}\\
\text{Twitter:} & a=1.90 \times 10^{-6}, \;\;\; b=1.54, \;\;\; c=2.35 \times 10^{-8}
\end{eqnarray}
}

\subsubsection*{Data and code availability}

All analysis code, as well as mention frequency data and supplementary data, have been deposited in GitHub (\url{https://github.com/epfl-dlab/post-mortem-memory}).

\section*{Acknowledgments}
R.W. was partly supported by Swiss National Science Foundation grant 200021\_185043, Collaborative Research on Science and Society (CROSS), and gifts from Google, Facebook, and Microsoft.
J.L. is a Chan Zuckerberg Biohub investigator and was partly supported by NSF grants OAC-1835598 (CINES), OAC-1934578 (HDR), CCF-1918940 (Expeditions), and IIS-2030477 (RAPID); Stanford Data Science Initiative; and Chan Zuckerberg Biohub.
We thank Janice Lan for help with taxonomy construction,
Spinn3r for data access,
and Ahmad Abu-Akel, Micha\l{} Kosi\'nski, and Michele Catasta for helpful discussions.

\renewcommand\refname{References}
\bibliographystyle{plain}
\bibliography{references,references_manual}

\end{document}
